\documentclass{article}
\usepackage{spconf,amsmath,graphicx}
\usepackage{cite}
\usepackage{amssymb,amsfonts}
\usepackage{algorithmic}
\usepackage{graphicx}
\usepackage{textcomp}
\usepackage{xcolor}
\usepackage{amssymb}
\usepackage{amsmath}
\usepackage{float}
\usepackage{subfig}
\usepackage{url}
\usepackage{hanging}

\title{SUBJECTIVE QUALITY EVALUATION OF POINT CLOUDS USING A\\HEAD MOUNTED DISPLAY}
\name{Joao Prazeres, Rafael Rodrigues, Manuela Pereira, Antonio M. G. Pinheiro\thanks{Research funded by the Portuguese FCT-Fundacao para a Ciencia e Tecnologia  under the project UIDB/50008/2020, PLive  X-0017-LX-20, and by operation Centro-01-0145-FEDER-000019 - C4 - Centro de Competencias em Cloud Computing.}}
\address{Universidade da Beira Interior \& Instituto de Telecomunicacoes, Covilha, Portugal}
\begin{document}
%
\maketitle
\begin{abstract}
This paper reports on a subjective quality evaluation of static point clouds encoded with the MPEG codecs V-PCC and G-PCC, the deep learning-based codec RS-DLPCC, and the popular Draco codec. 18 subjects visualized 3D representations of distorted point clouds using a Head Mounted Display, which allowed for a direct comparison with their reference. The Mean Opinion Scores (MOS) obtained in this subjective evaluation were compared with the MOS from two previous studies, where the same content was visualized either on a 2D display or a 3D stereoscopic display, through the Pearson Correlation, Spearman Rank Order Correlation, Root Mean Square Error, and the Outlier Ratio. The results indicate that the three studies are highly correlated with one another. Moreover, a statistical analysis between all evaluations showed no significant differences between them.
\end{abstract}
\section{Introduction}
Recently, point clouds have emerged as a popular method of 3D representation. 
Point clouds consist of a set of points represented in the 3D space, typically by their cartesian coordinates ($x,y,z$), with associated attributes, such as RGB, reflectance values, normal vectors, or information from physical sensors. These allow for an accurate representation of several types of 3D content, useful in a wide range of applications, such as virtual (VR), augmented (AR) and mixed reality applications, 3D printing, automation and robotics, computer graphics and gaming, or medical applications, among others. 
However, point clouds usually contain a large amount of information. Hence, there is a need for efficient algorithms to compress point cloud data, as well as suitable quality models to evaluate the compression performance.

Two of the most well-know point cloud compression solutions are the MPEG codecs V-PCC~\cite{V-PCC} 
(Video Point Cloud Compression) and 
G-PCC~\cite{G-PCC} (Geometry Point Cloud Compression). 
Inspired by the success of deep learning-based image coding, several other solutions have been proposed recently ~\cite{GuardaADLPCC,Jianqiang-PCGCv2,quach2020improved}, and studied regarding their compression quality~\cite{Prazeresacm22} and performance stability over different training sessions~\cite{Prazereseuvip22}. The Google codec Draco has also been studied in~\cite{EI2022}.

Multiple recent works have focused on quality evaluation methodologies for point clouds, with the most relevant being based on subjective evaluation.
In~\cite{8463406} and \cite{surf}, the authors established quality models for geometry-only point clouds. Compression artifacts using prior encoding schemes are evaluated in~\cite{8743258,8122239,Perry2019a}, whereas a wide range of high-performance point cloud codecs have been previously studied 
\cite{AlexiouMMSP2017,8803298,ICIP2020,PerryEuvip2022}.
In \cite{surf}, the authors concluded that using a 3D visualization with surface reconstruction or a 2D visualization did not change the evaluation. Alexiou \textit{et al.} \cite{AlexiouMMSP2017} used an AR environment for subjective quality, and Subramanyam \textit{et al.} \cite{ViolaVr} used VR to evaluate dynamic point clouds, with two degrees of freedom. Research on environments for subjective evaluation has also been previously conducted \cite{Alexiou:277378}. 

In this work, a subjective evaluation using a Head Mounted Display (HMD) is reported, and the results compared with previous studies using a 2D display \cite{EI2022} and 3D stereoscopic visualization~\cite{ICIP2022}.
Thus, the same codecs were tested, i.e., the MPEG codecs V-PCC and G-PCC, RS-DLPCC~\cite{GuardaRS-DLPCC} and Draco, for a direct  comparison.

Using HMDs in subjective quality assessment bears several advantages. From a test preparation standpoint, typical point cloud subjective evaluations using 2D displays require rendering a large amount of uncompressed high definition videos - preferably Full 4K - of the point clouds rotating around a selected axis. Uncompressed videos is a most, otherwise video compression artifacts could influence the perceived quality. This may result in videos with more than 400GB. HMDs allow using a direct representation of the point cloud, thus bypassing the need for the rendering of video sequences.

The remainder of this paper describes the evaluation methodology first, followed by an analysis of the obtained results and the drawn conclusions.

\begin{figure*}[t!]
    \centering
    \subfloat[\textit{Longdress}]{\label{fig:Long}\includegraphics[width=0.15\linewidth]{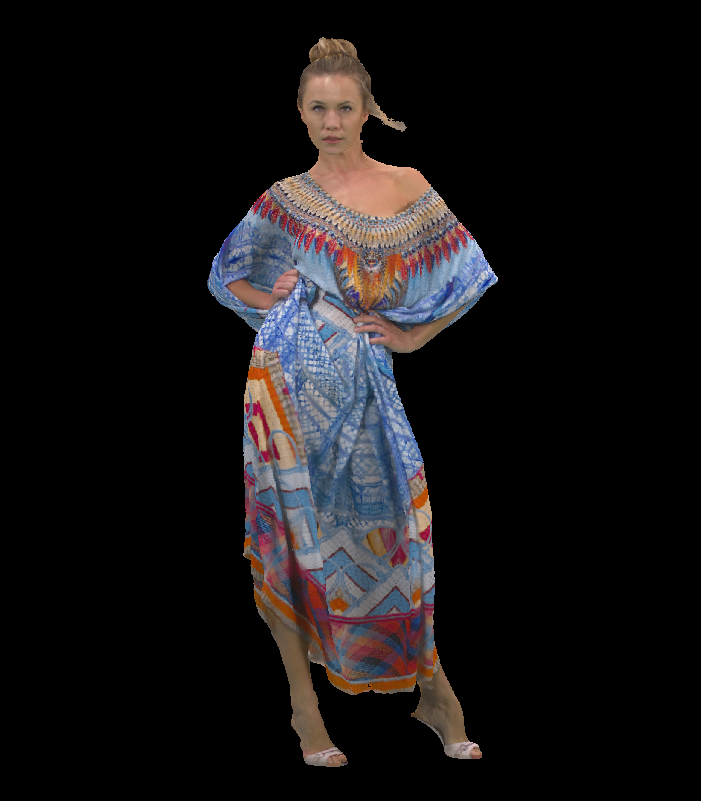}}\hfill
    \subfloat[\textit{Soldier}]{\label{fig:Soldier}\includegraphics[width=0.15\linewidth]{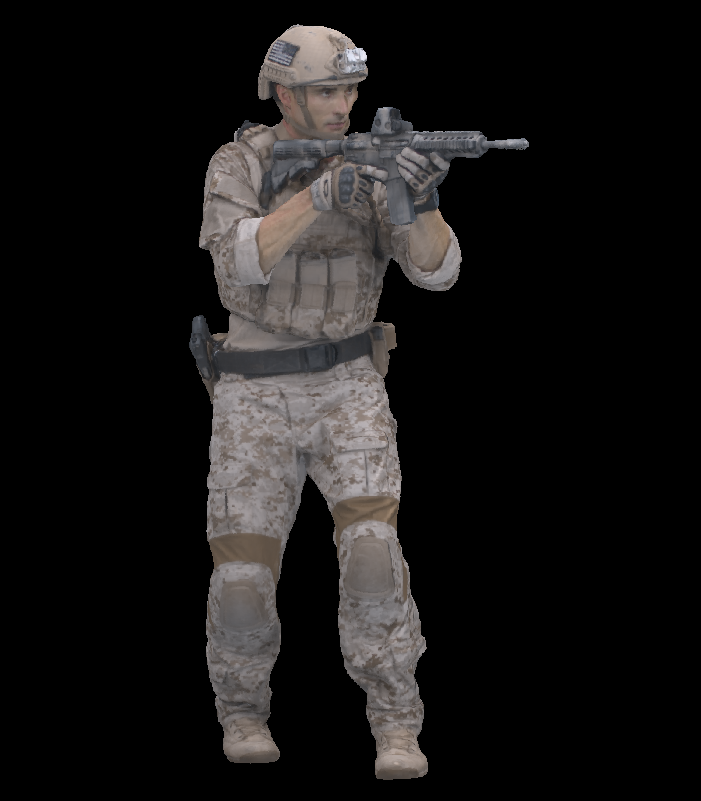}}\hfill
    \subfloat[\textit{Rhetorician}]{\label{fig:Rhe}\includegraphics[width=0.15\linewidth]{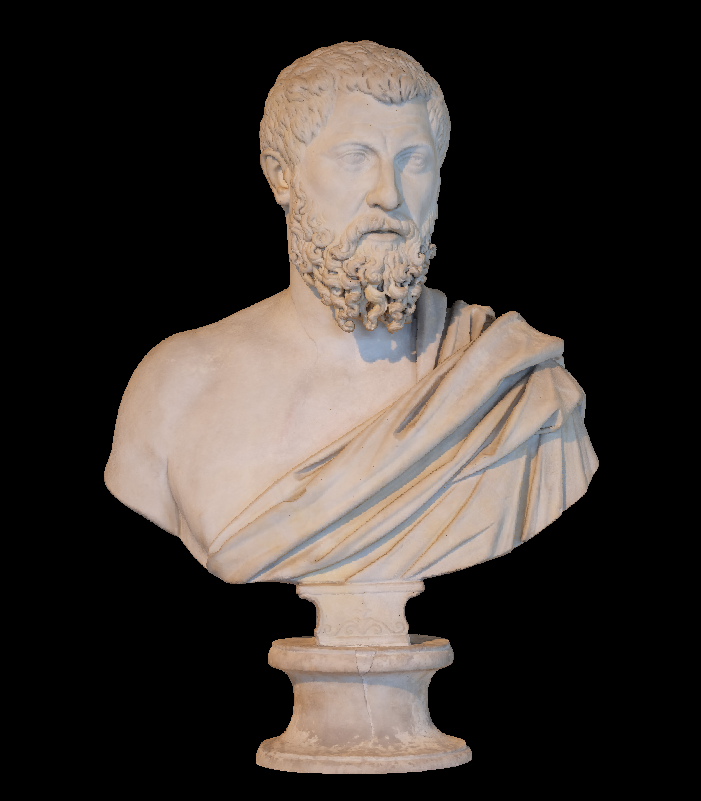}}\hfill
    \subfloat[\textit{Guanyin}]{\label{fig:Guan}\includegraphics[width=0.15\linewidth]{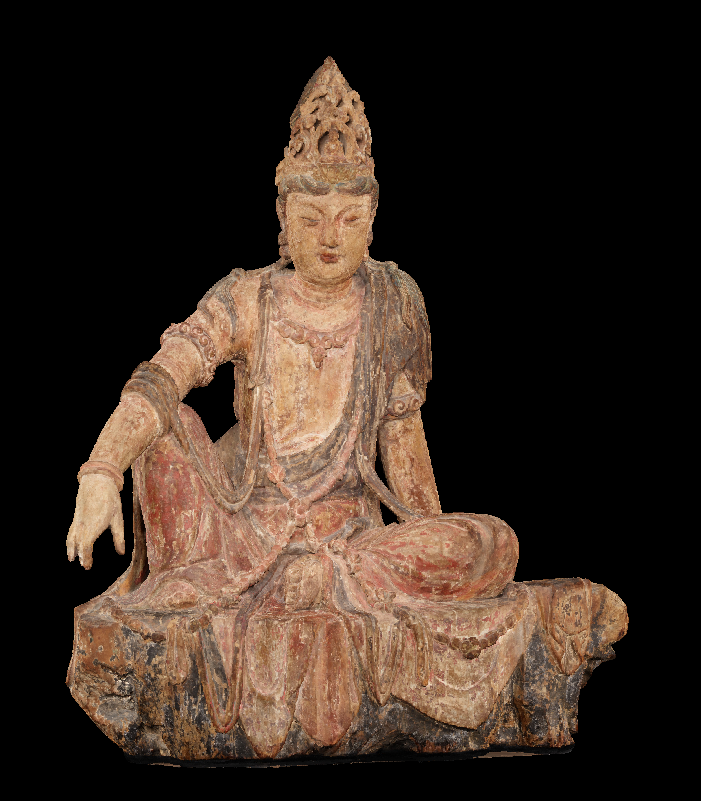}}\hfill
    \subfloat[\textit{Romanoillamp}]{\label{fig:Roman}\includegraphics[width=0.15\linewidth]{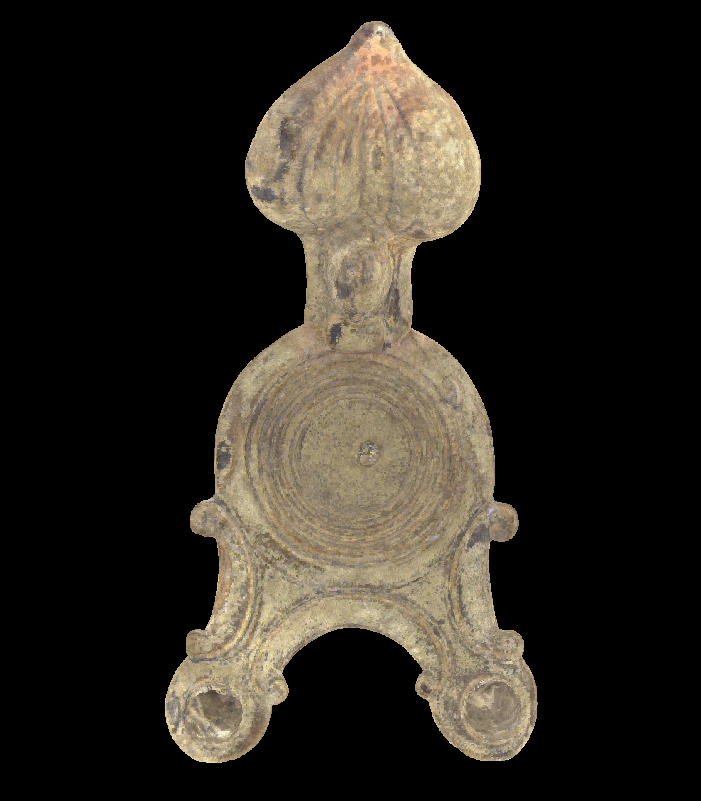}}\hfill
    \subfloat[\textit{Bumbameuboi}]{\label{fig:Bumba}\includegraphics[width=0.15\linewidth]{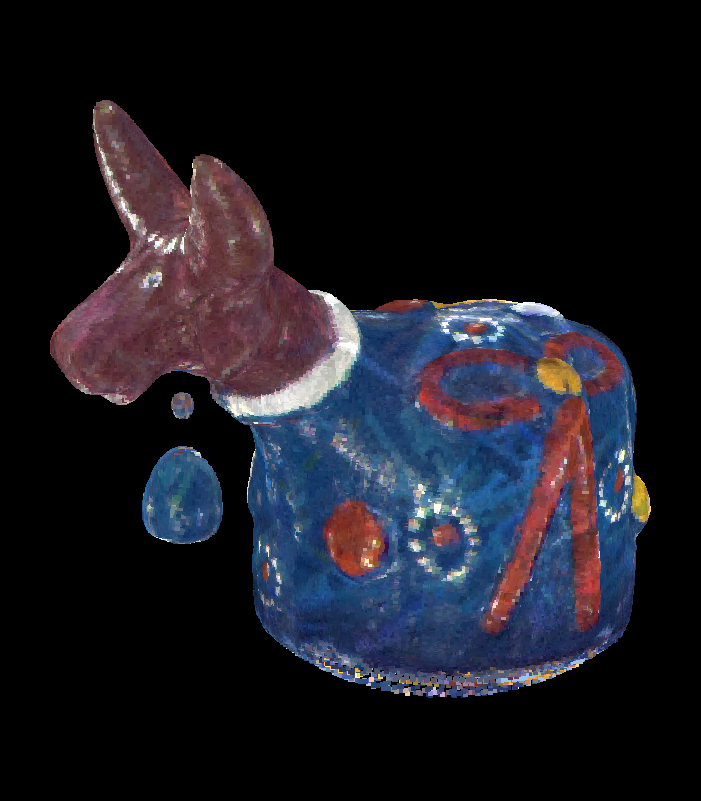}}\\
    \caption{Point Cloud testing set.}
    \label{PCS}
\end{figure*}  

\section{Evaluation Methodology}
\vspace{-0.3cm}
\subsection{Point Cloud Data Selection}
For this study, a set of six point clouds was selected (Fig.~\ref{PCS}), all containing geometry and texture information. The set consisted of frames 1300  from the \textit{Longdress} and 690 from the \textit{Soldier} dynamic point clouds\footnote{https://jpeg.org/plenodb/},  the static point clouds \textit{Rhetorician} and \textit{Guanyin}, from the EPFL dataset, and \textit{Romanoillamp} and \textit{Bumbameuboi}, from the University of Sao Paulo Point Cloud Dataset\footnote{http://uspaulopc.di.ubi.pt}. The first two represent a human figure, and the last four represent cultural heritage, providing diversity of geometrical, textural, and point density characteristics within the dataset~\cite{ICIP2022}.
\vspace{-0.3cm}

\begin{figure}[t!]
    \centering
     \vspace{-0.7cm}
    \subfloat[\textit{Longdress}]{\label{fig:SoldierMos}\includegraphics[width=0.50\linewidth]{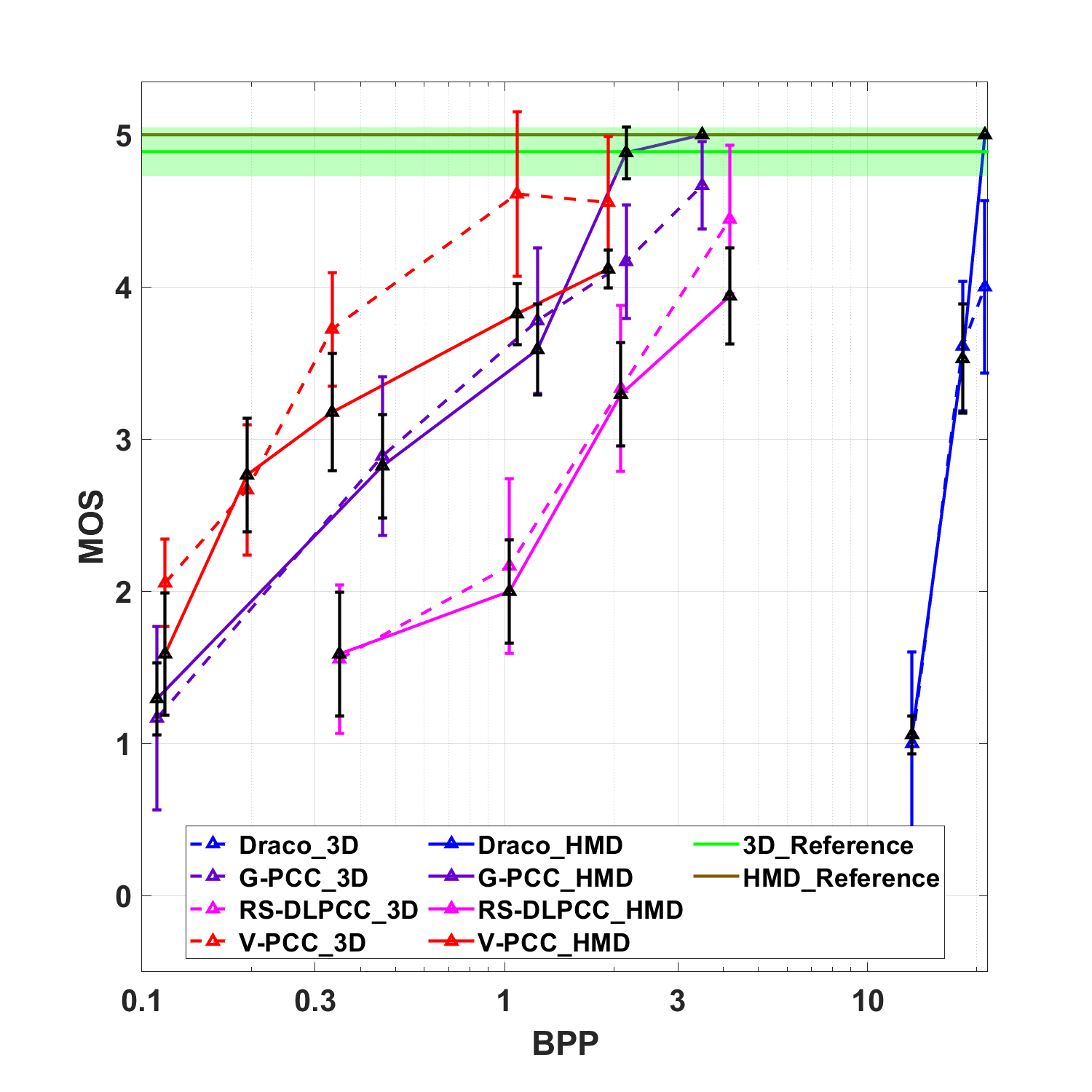}}\hfill
    \subfloat[\textit{Soldier}]{\label{fig:GuanMos}\includegraphics[width=0.50\linewidth]{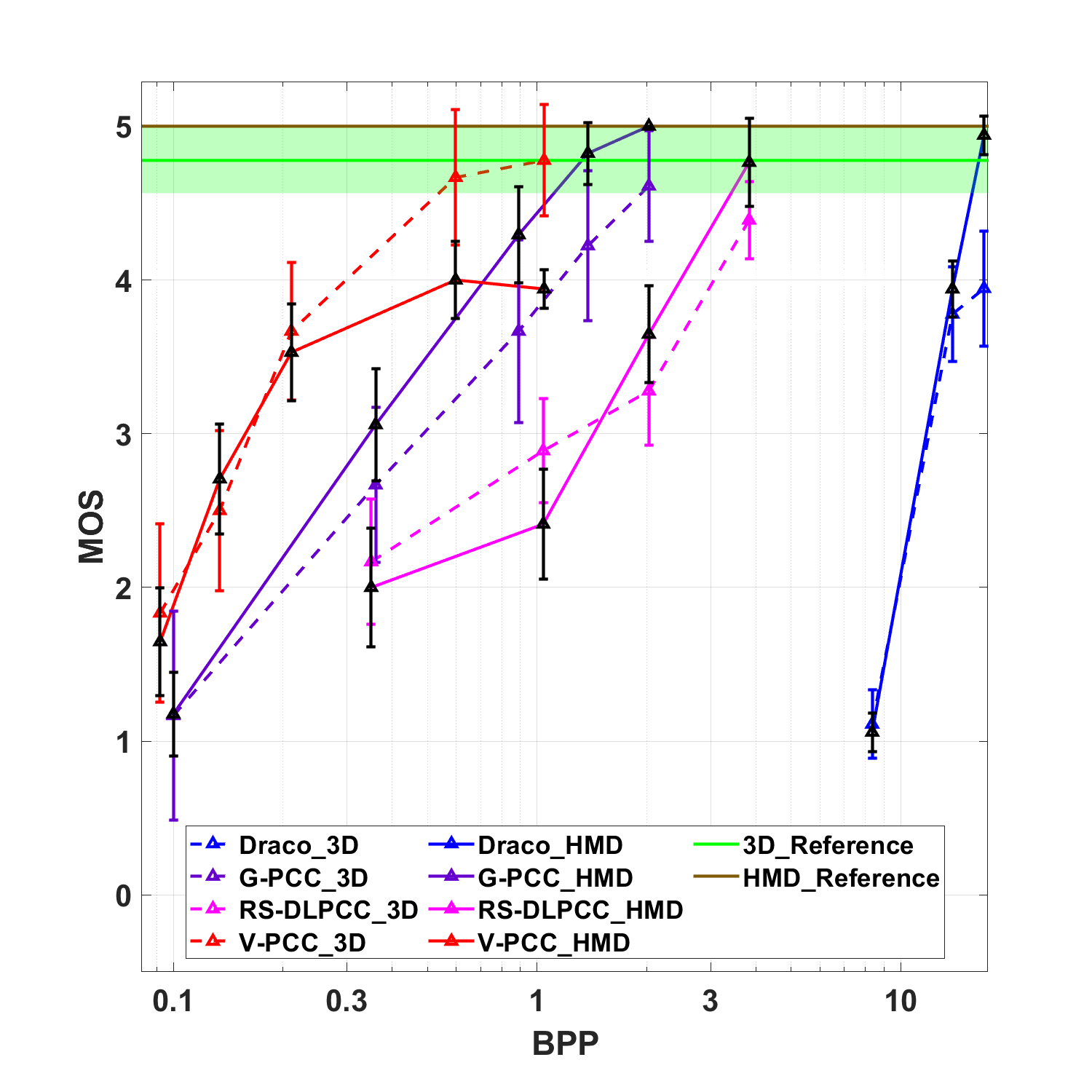}}\vspace{-0.4cm}
    \subfloat[\textit{Rhetorician}]{\label{fig:LongMos}\includegraphics[width=0.50\linewidth]{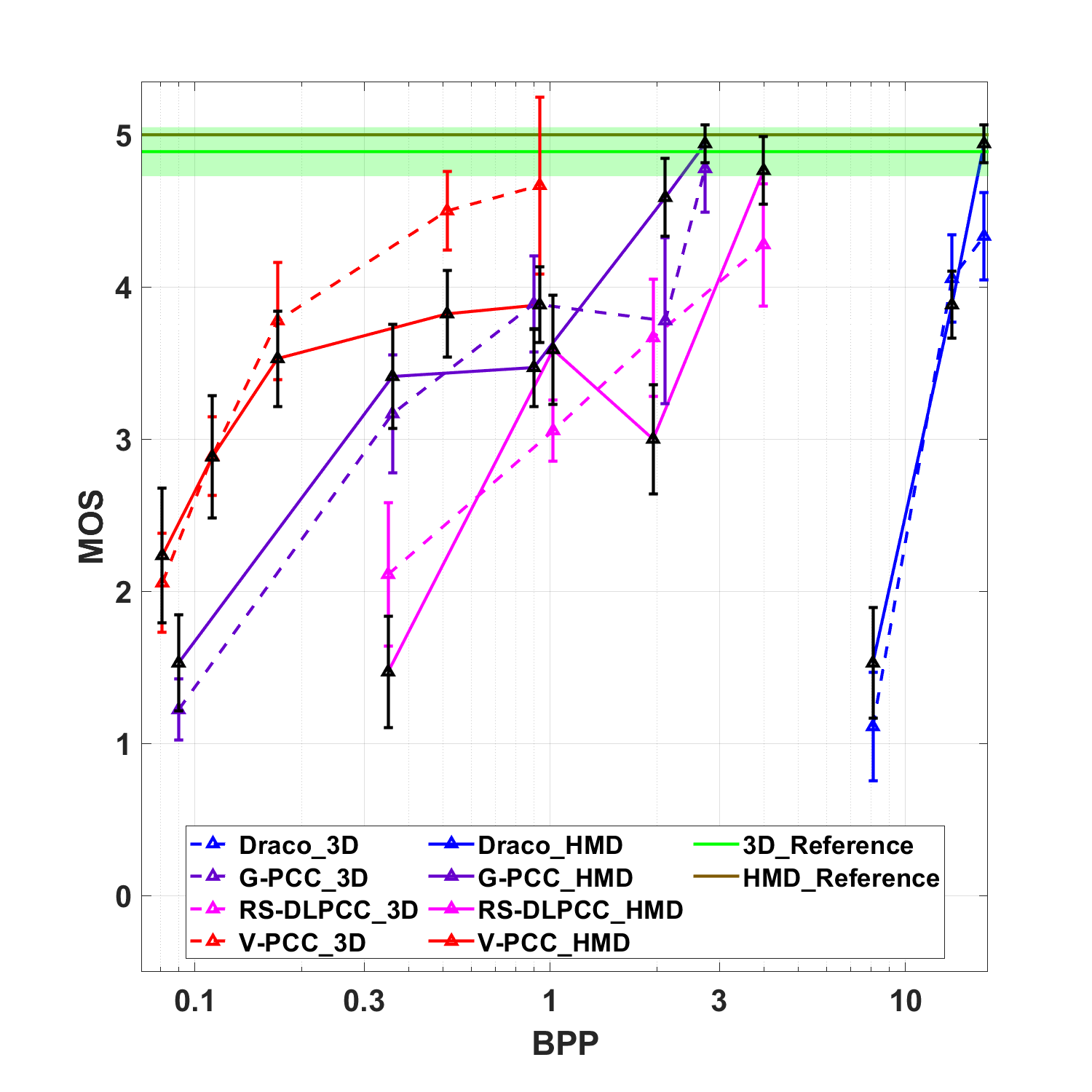}}\hfill
    \subfloat[\textit{Guanyin}]{\label{fig:RheMos}\includegraphics[width=0.50\linewidth]{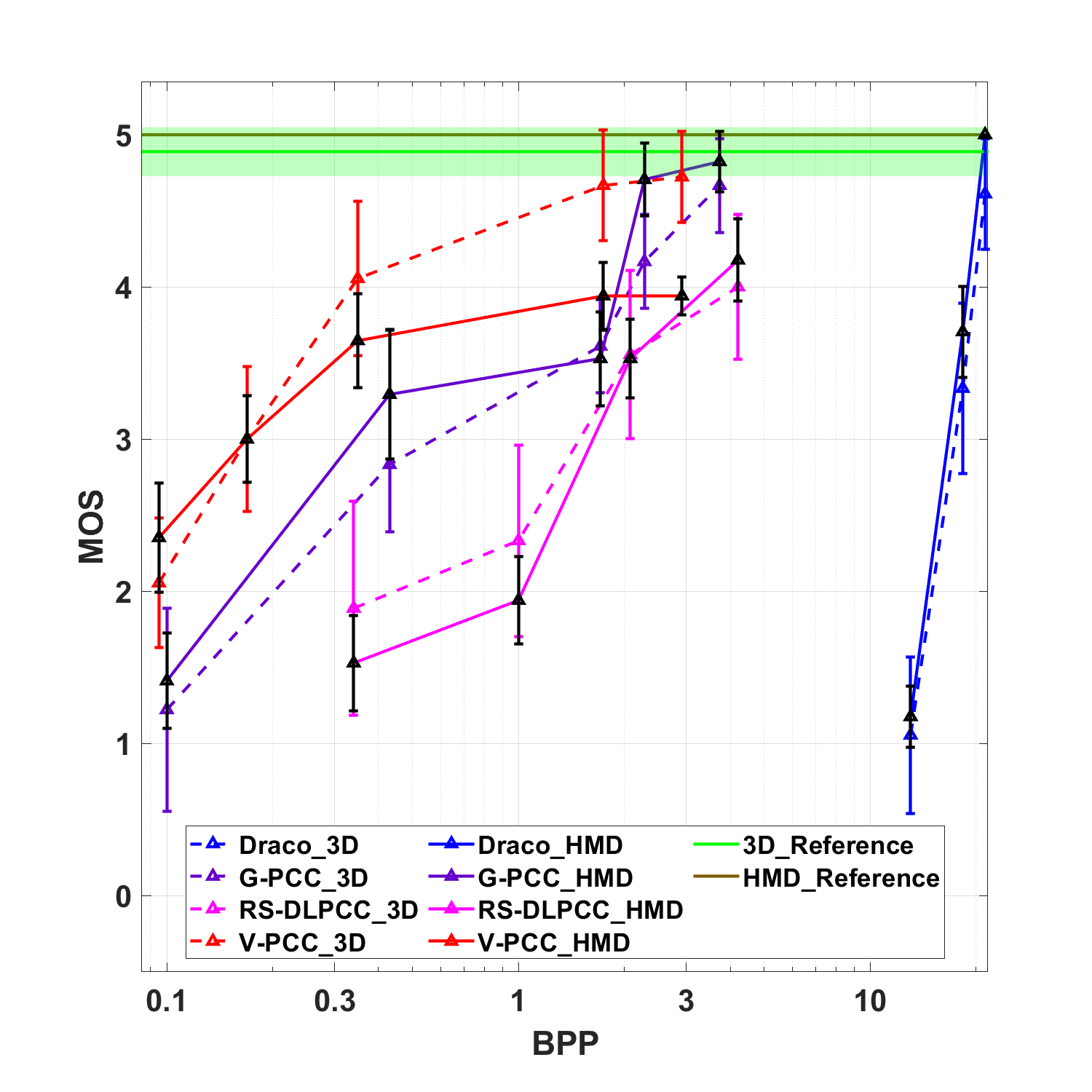}}\vspace{-0.4cm}
    \subfloat[\textit{Romanoillamp}]{\label{fig:RomanMos}\includegraphics[width=0.50\linewidth]{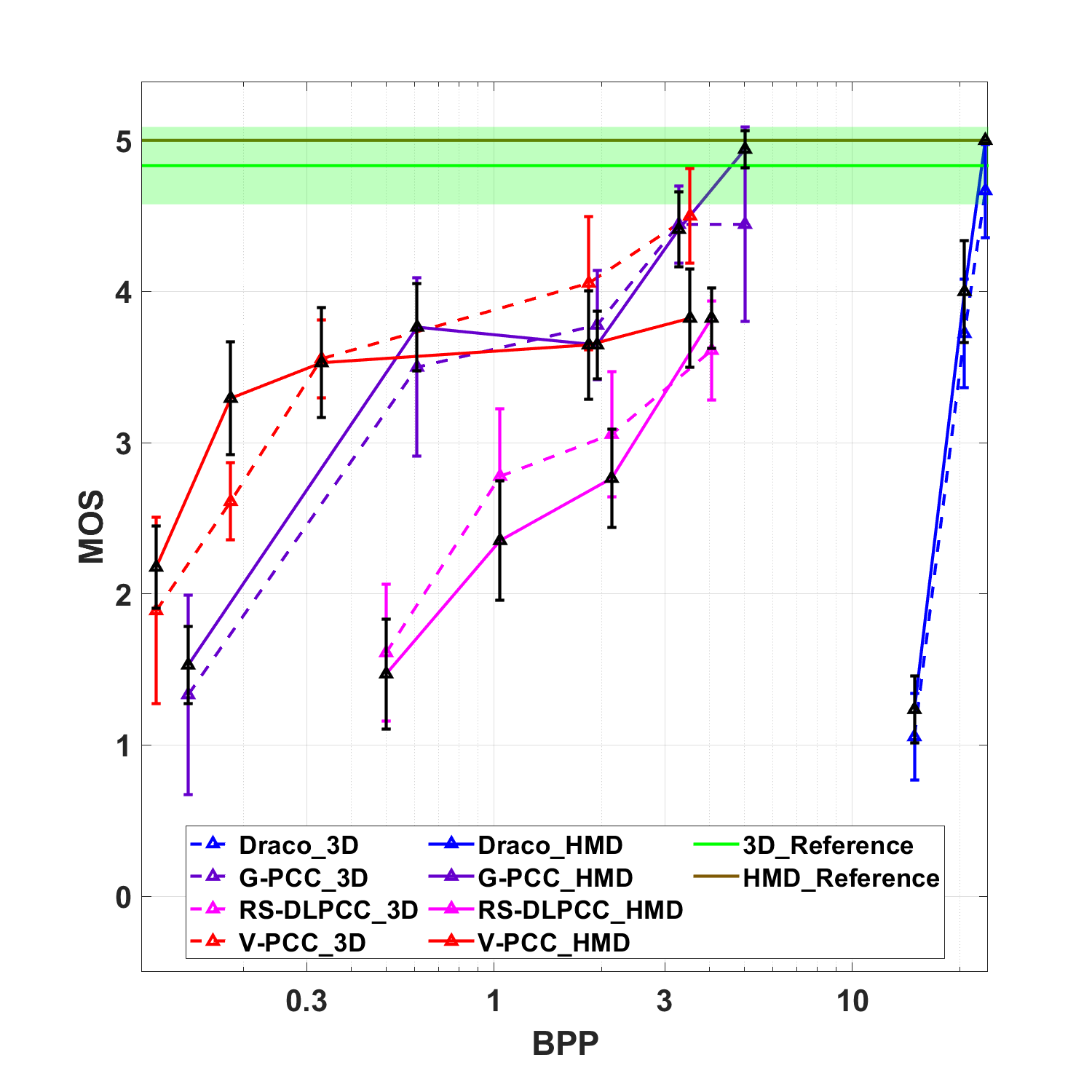}}\hfill
    \subfloat[\textit{Bumbameuboi}]{\label{fig:BumbaMos}\includegraphics[width=0.50\linewidth]{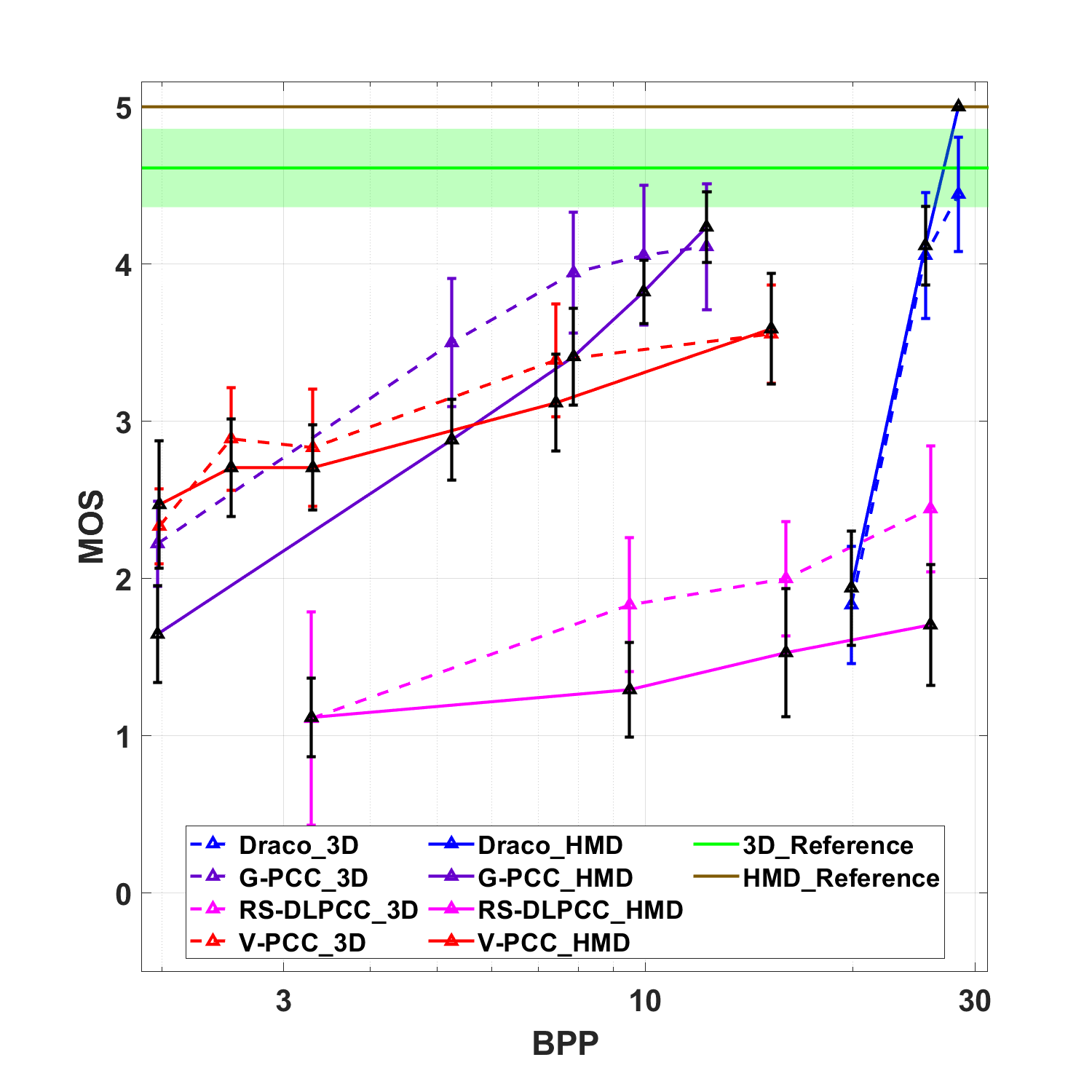}}\\
    \caption{MOS vs. bpp with 95\% CIs, considering a Gaussian distribution.}
    \label{MOSBPP}
\end{figure}

\subsection{Selected Codecs}

A short description of the used codecs can be found in \cite{EI2022,ICIP2022}. The MPEG V-PCC~\cite{V-PCC} (using HEVC as video codec) and G-PCC (using Octree for geometry and Prediction-plus-Lifting for texture)~\cite{G-PCC} codecs were selected as representative of two state of the art solutions. 
RS-DLPCC \cite{GuardaRS-DLPCC} is a deep learning solution with added scalability property, and Draco was chosen because of its popularity.
The codecs parameters are listed in tables \ref{table:mpegex1} and \ref{table:dracoParams}.




\begin{table}[t!]
\begin{center}
\caption{Coding parameters for G-PCC and V-PCC.}\label{G-PCC Parameters}%
\label{table:mpegex1}
\large
\resizebox{\linewidth}{!}{%
\begin{tabular}{@{}|c|c|c|c|c|c|c|c|c|c|c|c|@{}}
\hline
\multicolumn{6}{|c|}{\textbf{G-PCC}} & \multicolumn{6}{|c|}{\textbf{V-PCC}}\\ \hline
\textbf{Rate} & R01 & R02 & R03 & R04 & R05 & \textbf{Rate} & R01 & R02 & R03 & R04 & R05\\\hline
\textbf{QP} &  46 &  40 &  34 &  28 & 22 & \textbf{Geometry QP} & 36 & 32 & 28 & 20 & 16\\\hline 
\textbf{pQS} & 0.25 & 0.5 & 0.75 & 0.875 & 0.9375 & \textbf{Texture QP} & 47 & 42 & 37 & 27 & 22\\\hline
\multicolumn{6}{|c|}{} &Occupancy Map & \multicolumn{4}{|c|}{4} & 2\\\hline
\end{tabular}%
}
\end{center}
\vspace{-0.5cm}
\end{table}

\begin{table}[t!]
    \centering
    \footnotesize
    \caption{QP for the Draco codec.}
    \begin{tabular}{|c|c|c|c|}
        \hline
        \textbf{Rate} & R01 & R03 & R05 \\ \hline
        \textbf{QP} & 7 & 9 & 10 \\\hline
        \end{tabular}
    \label{table:dracoParams}
\end{table}

\begin{table}[t!]
\begin{center}

\caption{Point size for each content.}
\centering
\Large
\resizebox{\linewidth}{!}{%
\begin{tabular}{@{}|c|c|c|c|c|c|c|c|c|c|c|c|@{}}
\hline
&\multicolumn{5}{|c|}{\textbf{V-PCC}} & \multicolumn{5}{c|}{\textbf{G-PCC}}\\\hline
\textbf{Content} & R01 & R02 & R03 & R04 & R05 & R01 & R02 & R03 & R04 & R05\\\hline
\textit{Bumbameuboi} & \multicolumn{5}{|c|}{0.008} & 0.012 & \multicolumn{4}{|c|}{0.01}\\\hline
\textit{Guanyin} & \multicolumn{5}{|c|}{0.002} & 0.006 & 0.004 &\multicolumn{3}{|c|}{0.002} \\\hline
\textit{Longdress} & \multicolumn{5}{|c|}{0.002} & 0.007 & 0.003 & \multicolumn{3}{|c|}{0.002} \\\hline
\textit{Rhetorician} & \multicolumn{5}{|c|}{0.002} & 0.007 & 0.004 & \multicolumn{3}{|c|}{0.002}  \\\hline
\textit{Romanoillamp} & \multicolumn{5}{|c|}{0.002}  & 0.006 & 0.003 & \multicolumn{3}{|c|}{0.002} \\\hline
Soldier &\multicolumn{5}{|c|}{0.002}& 0.007 & 0.004 & 0.003 & \multicolumn{2}{|c|}{0.002} \\\hline
&\multicolumn{5}{|c|}{\textbf{RS-DLPCC}} & \multicolumn{5}{c|}{\textbf{Draco}} \\\hline
\textit{Bumbameuboi} &-& 0.03 & 0.019 & 0.011 & 0.01 & 0.14 & - & 0.1 & - & 0.1 \\\hline
\textit{Guanyin} & -& 0.04 & \multicolumn{3}{|c|}{0.002} & 0.012 & - & 0.003 & - & 0.002  \\\hline
\textit{Longdress} & -& 0.04 & \multicolumn{3}{|c|}{0.002} & 0.013 & - & 0.004 & - & 0.002 \\\hline
\textit{Rhetorician} & -&  0.004 & 0.003 & \multicolumn{2}{|c|}{0.002} & 0.013 & - & 0.004 & - & 0.002 \\\hline
\textit{Romanoillamp} & -&  0.004 & 0.003 & \multicolumn{2}{|c|}{0.002}& 0.01 & - & 0.003 & - & 0.002 \\\hline
Soldier & -& 0.004 & \multicolumn{3}{|c|}{0.002} &  0.012 & - & 0.003 & - & 0.002 \\\hline

\end{tabular}%
\label{table:pointsize}}
\end{center}
\end{table}

\subsection{Data Generation and Experimental Setup}
The Unity\footnote{https://unity.com} software was used with the Pcx point cloud importer library\footnote{https://github.com/keijiro/Pcx}, which allows to manipulate and visualize point cloud data. The point size for each point cloud used in the subjective evaluation was adjusted as shown in Table~\ref{table:pointsize}, in order to  create continuous surfaces, thus avoiding perceptual effects caused by transparency~\cite{8463406,8743258}. 

The point clouds were positioned at a distance that would not cause discomfort to the subjects, which were seated in a fixed position. During the visualization stage, for each reference-distorted pair, only one version was visible at a time, starting always with a frontal view of the reference point cloud. Subjects were able to freely rotate the point cloud clockwise around the vertical axis, as well as alternating between the reference and the distorted point cloud at any time. Accessing the evaluation screen was only allowed after at least a full rotation of the point cloud and 6 commutations between the reference and the distorted point cloud. Each test sequence was unique and randomized, while ensuring that the same content was never shown twice in a row. Hidden reference-reference pairs were also included, 
resulting in a total of 108 pairs.

\begin{figure}[t!]
\centering
\vspace{-0.6cm}
\subfloat[2D vs. HMD]{\includegraphics[width=0.48\linewidth]{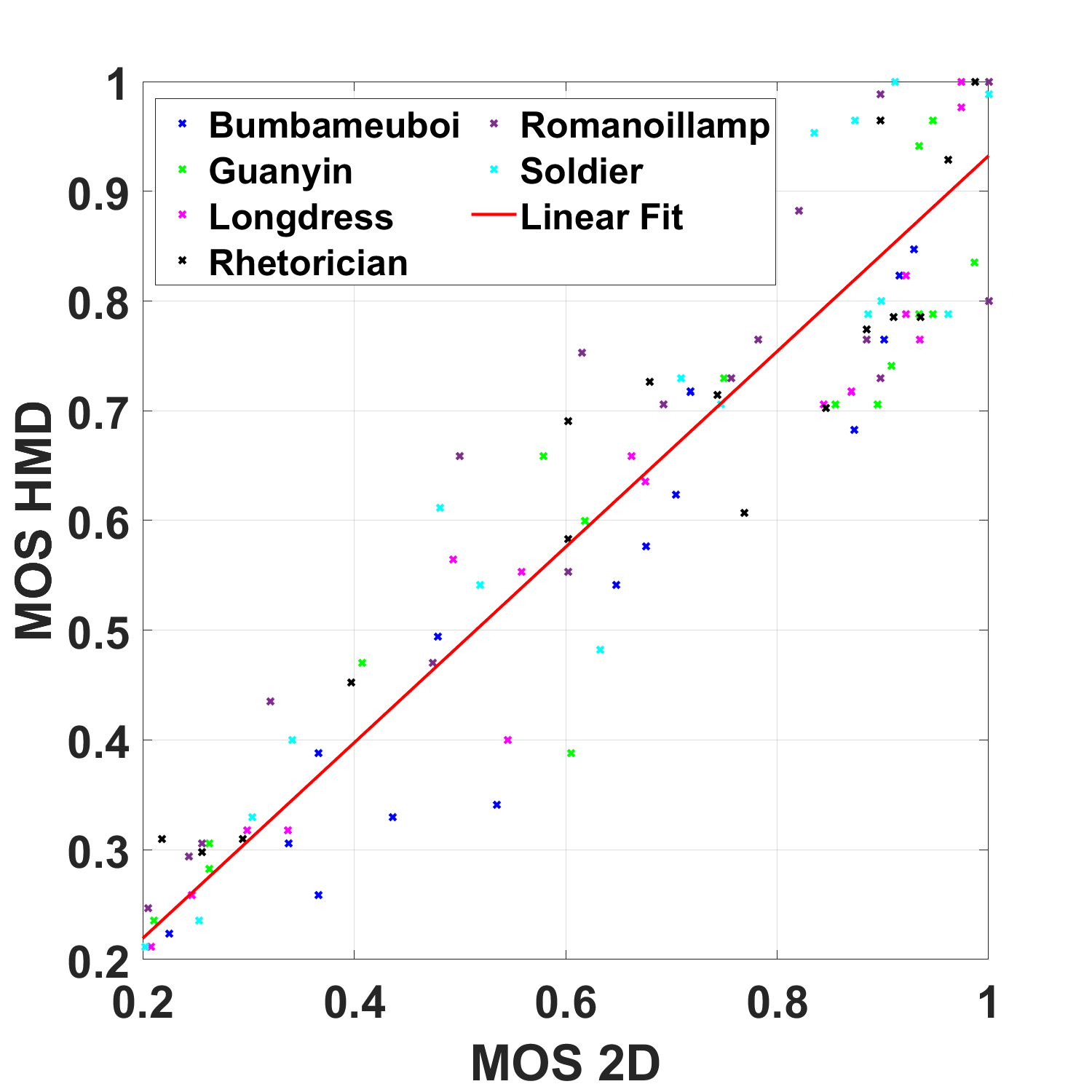}}
\subfloat[3D vs. HMD]{\includegraphics[width=0.48\linewidth]{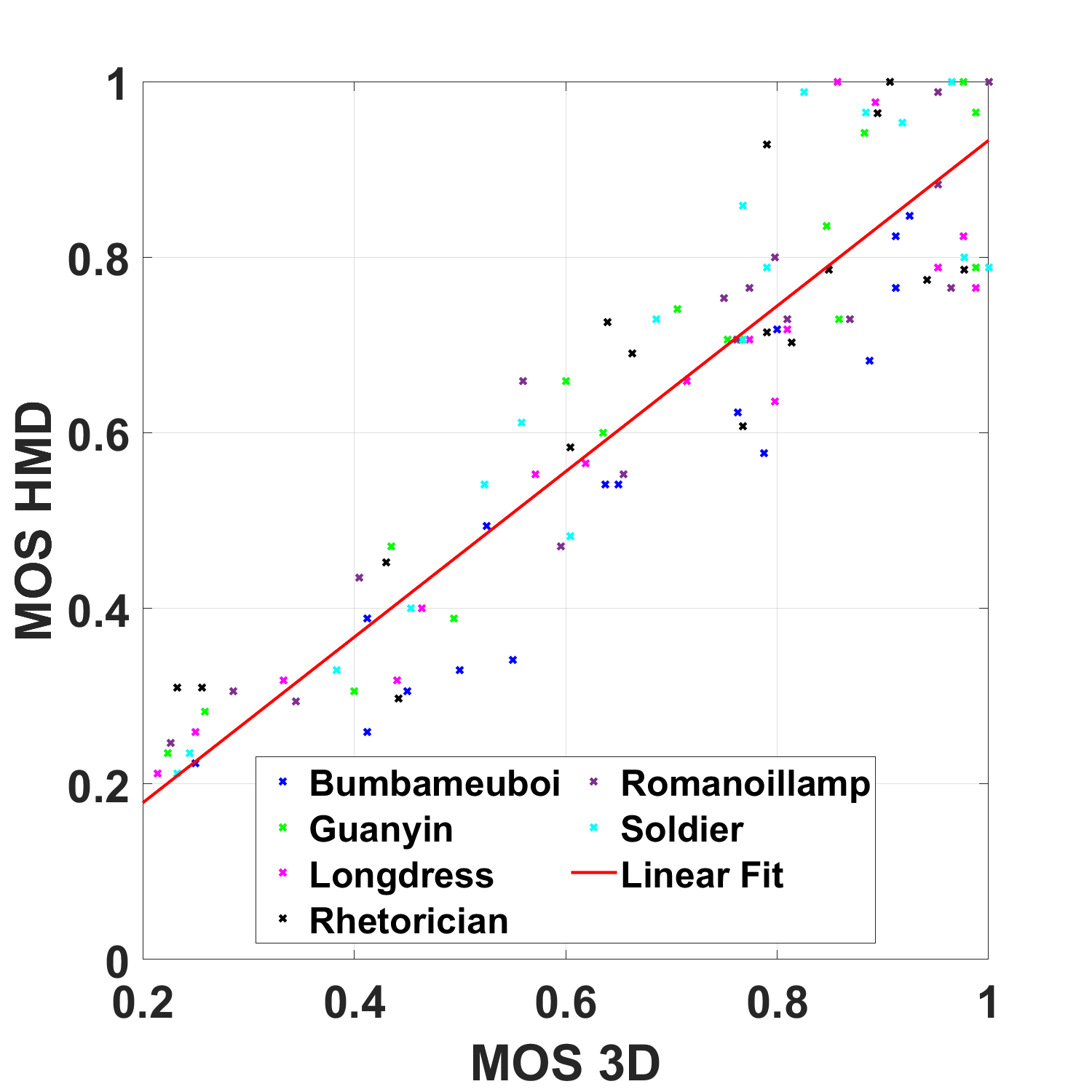}}
\caption{Comparison between the MOS obtained using the 2D and the 3D representations, 
and respective linear fitting.}
\label{fig:MOS3DMOS2D}
\end{figure}

\begin{table}[t!]
    \centering
    \caption{Correlation statistics for the 2D~\cite{EI2022} vs. HMD and 3D~\cite{ICIP2022} vs. HMD comparisons.}
    \footnotesize
    \begin{tabular}{|c|c|c|c|c|}
     \hline
        \textbf{Test} & \textbf{PCC} & \textbf{SROCC} & \textbf{RMSE} & \textbf{OR} \\ \hline
            2D VS HMD & 0.943 & 0.942 & 0.109 & 0.539 \\ \hline
            3D VS HMD & 0.934 & 0.924 & 0.118 & 0.588 \\ \hline
    \end{tabular}
    \label{table:testCorrelation}
 \end{table}


\begin{table}[!t]
    \centering
      \caption{Kruskal-Wallis \textit{p}-values\cite{KruskalWallis} for the 2D~\cite{EI2022} vs. HMD and 3D~\cite{ICIP2022} vs. HMD comparisons.}\label{tab:statDif}
      \resizebox{\linewidth}{!}{%
    \begin{tabular}{|c|c|c|c|c|c|}
    \hline
      \textbf{Test} & \textbf{Global} & \textbf{Draco} & \textbf{RS-DLPCC} & \textbf{G-PCC} & \textbf{V-PCC}  \\ \hline
        2D vs HMD & 0.818 & 0.680 & 0.885 & 0.544 & 0.657 \\ \hline
        3D vs HMD & 0.881 & 0.289 & 0.433 & 0.506 & 0.318 \\ \hline
    \end{tabular}%
    }
\end{table}
\begin{table*}[t!]
\begin{center}
\vspace{-0.2cm}
\caption{Average of the 95\% CIs for the three conducted experiments, considering a Gaussian distribution.}
\centering
\resizebox{\linewidth}{!}{%
\begin{tabular}{@{}|c|c|c|c|c|c|c|c|c|c|c|c|c|c|c|c|c|c|c|c|c|c|c|c|c|@{}}
\hline
\multicolumn{22}{|c|}{\textbf{3D\cite{ICIP2022} vs HMD}}\\ \hline
&&\multicolumn{5}{|c|}{\textbf{V-PCC}} & \multicolumn{5}{c|}{\textbf{G-PCC}} &\multicolumn{5}{|c|}{\textbf{RS-DLPCC}} & \multicolumn{5}{c|}{\textbf{Draco}} \\\hline
\textbf{Test}& \textbf{Global} & R01 & R02 & R03 & R04 & R05 & R01 & R02 & R03 & R04 & R05 & - & R02 & R03 & R04 & R05 & R01 & - & R03 & -  & R05\\\hline
\textbf{3D} &\textbf{0.447} &0.441 & 0.404 & 0.417 & 0.428 & 0.411 & 0.549 & 0.509 & 0.436 & 0.431& 0.407  &-&0.571& 0.466 & 0.466 & 0.418 & 0.420&- & 0.417&- & 0.403\\ \hline
\textbf{HMD}&\textbf{0.291} & 0.397 & 0.375 & 0.348 & 0.428 & 0.232 & 0.304 & 0.359 & 0.436 & 0.235 & 0.120&-& 0.371 & 0.363 & 0.357 & 0.300 & 0.249 & - &0.294 & - & 0.044 \\ \hline



\hline
\multicolumn{22}{|c|}{\textbf{2D\cite{EI2022} vs HMD (considering the first 16 subjects)}}\\ \hline
&&\multicolumn{5}{|c|}{\textbf{V-PCC}} & \multicolumn{5}{c|}{\textbf{G-PCC}} &\multicolumn{5}{|c|}{\textbf{RS-DLPCC}} & \multicolumn{5}{c|}{\textbf{Draco}} \\\hline
\textbf{Test}& \textbf{Global} & R01 & R02 & R03 & R04 & R05 & R01 & R02 & R03 & R04 & R05 & - & R02 & R03 & R04 & R05 & R01 & - & R03 & -  & R05\\\hline
\textbf{2D} &\textbf{0.335}  & 0.396 & 0.421 & 0.378 & 0.339 & 0.401 & 0.307 & 0.409 & 0.357 & 0.355 & 0.294 & - & 0.245 & 0.393 & 0.369 & 0.350 & 0.094 & - & 0.351 & - & 0.2413\\ \hline
\textbf{HMD}&\textbf{0.296} &0.418 & 0.378 & 0.310 & 0.339 & 0.241 & 0.320 & 0.346 & 0.357 & 0.246 & 0.128 & - & 0.389 & 0.366 & 0.371 & 0.315 & 0.261 &- & 0.290 &-& 0.048 \\ \hline

\end{tabular}%
\label{table:confidenceIntervals}}
\end{center}
\end{table*}

Before starting the evaluation, the subjects were instructed on how to use the HMD setup. Moreover, a set of five point clouds, depicting the \textit{Redandblack} point cloud (not included in the final test sequence) with five levels of degradation was also used to test the setup and the evaluation process. 
In fact, the training process was very similar to the conducted in the previous works \cite{EI2022,ICIP2022}, although in this case the HMD setup was used.

\begin{table}[t!]
\Large
\caption{Performance of the objective metrics using the 3D~\cite{ICIP2022} and HMD subjective scores as ground truth. The best metric is in bold, and the second best is in italic.}
\resizebox{\linewidth}{!}{%
\begin{tabular}{|l||c|c|c|c||c|c|c|c|}
\hline
& \multicolumn{4}{|c||}{\textbf{3D MOS}}
& \multicolumn{4}{|c|}{\textbf{HMD MOS}}\\ \hline
\textbf{\textit{Metric}} & \textbf{\textit{PCC}} & \textbf{\textit{SROCC}} & \textbf{\textit{RMSE}} & \textbf{\textit{OR}} &
 \textbf{\textit{PCC}} & \textbf{\textit{SROCC}} & \textbf{\textit{RMSE}} & \textbf{\textit{OR}} \\ \hline \hline
PSNR MSE D1 & 0.882 & 0.893 & 0.142 & 0.716 & \textit{0.881} & \textit{0.875} & \textit{0.144} & \textit{0.696} \\ \hline
PSNR MSE D2 & 0.855 & 0.851 & 0.157 & 0.657 & 0.826 & 0.842 & 0.172 & 0.637 \\ \hline
PointSSIM & 0.871 & 0.866 & 0.148 & 0.686 & 0.856 & 0.867 & 0.157 & 0.706 \\ \hline
1 - PCQM &\textbf{ 0.934} & \textbf{0.924} & \textbf{0.108} & \textbf{0.637} & \textbf{0.947} & \textbf{0.946} & \textbf{0.098} & \textbf{0.598} \\ \hline
GraphSIM & \textit{0.906} & \textit{0.872} & \textit{0.128} & \textit{0.598} & 0.876 & 0.890 & 0.147 & 0.706 \\ \hline

\end{tabular}%
}
\label{table:metricsResult}

\end{table}

\begin{figure}[t!]
\centering
\vspace{-0.4cm}
\subfloat[PSNR MSE D1]{\includegraphics[width=0.48\linewidth]{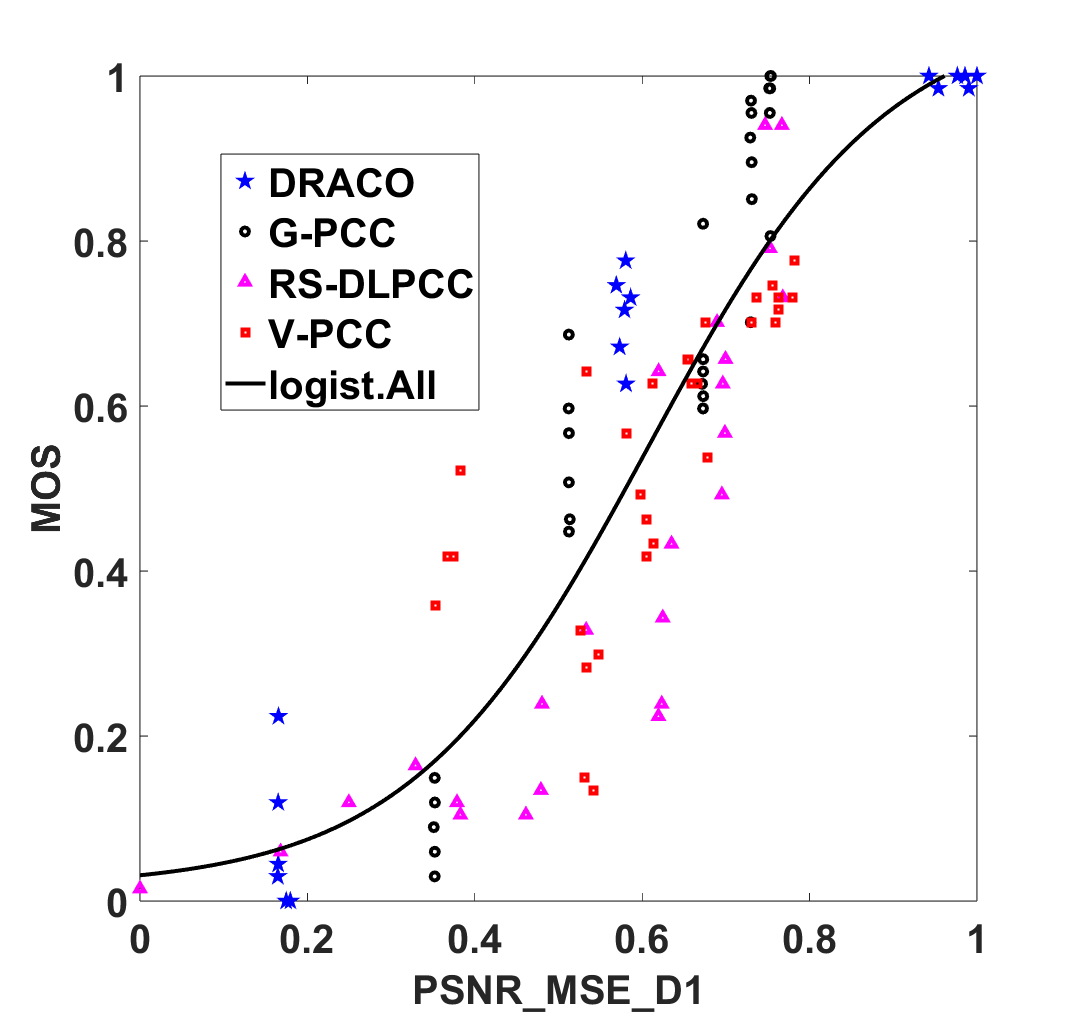}}
\subfloat[PCQM]{\includegraphics[width=0.48\linewidth]{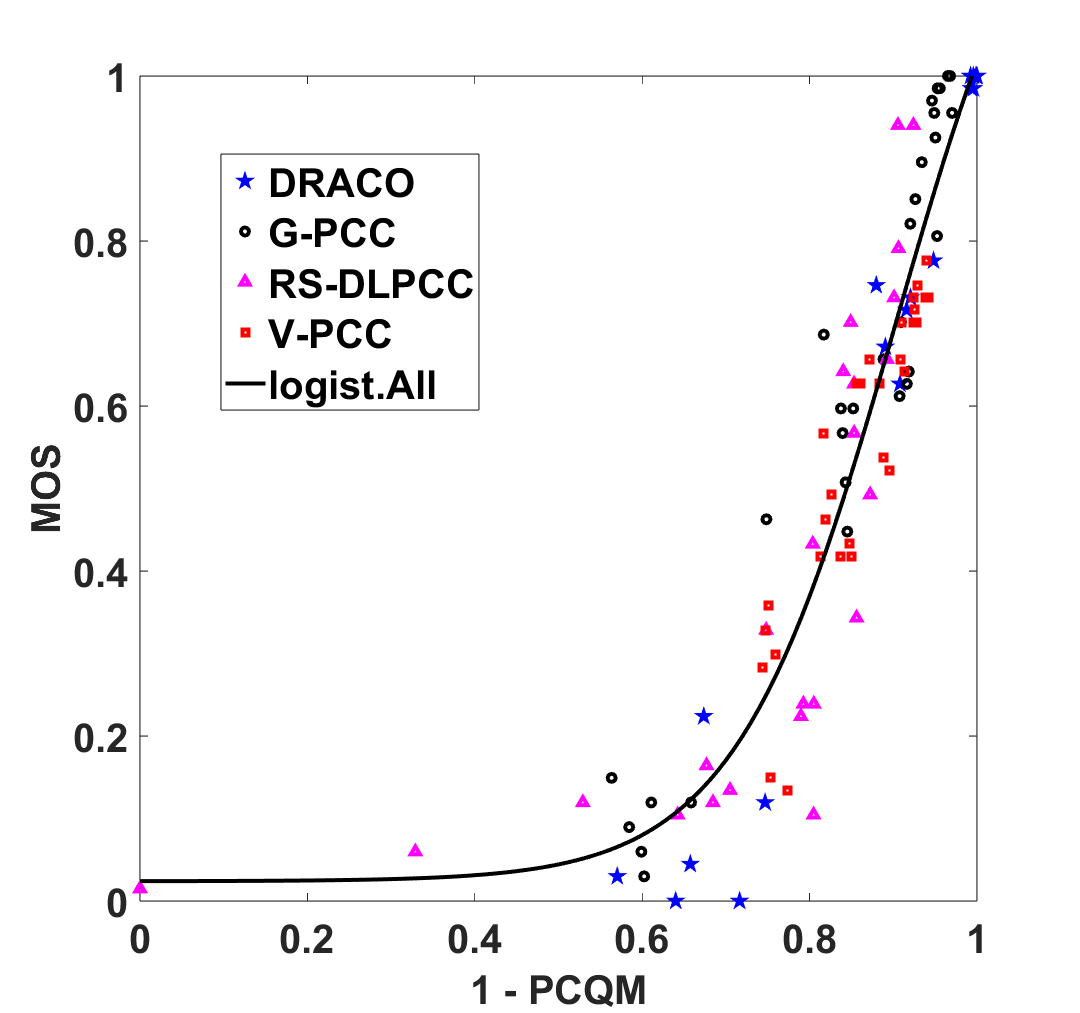}}
\caption{Relation between metrics and MOS, and Logistic fitting curve.}
\vspace{-0.5cm}
\label{fig:LogisticF}
\end{figure}

A Double Stimulus Impairment Scale method was used with a five-level rating scale (1 - very annoying, 2 - slightly annoying, 3 - annoying, 4 - perceptible, but not annoying, 5 - imperceptible)~\cite{ITU-T_BT500}. The experiment was conducted in the subjective test laboratory of the Image and Video Technology Group of Universidade da Beira Interior, using a HTC Vive Pro Headset, with a refresh rate of 90Hz, a field of view of 110$^{\circ}$, and resolution of $2880\times1600$ ($1440\times1600$ per eye). 
A total of 18 subjects participated in the experiment, 9 male and 9 female, with ages ranging between 20 and 32 years old (23.6$\pm$2.9). The Mean Opinion Score (MOS) for each test instance was computed by averaging the scores of all subjects for each test point cloud.

\vspace{-0.4cm}

\section{Results}
\subsection{Subjective Evaluation Results}

Fig. \ref{MOSBPP} shows the MOS obtained from the experiment described in this paper and the experiment conducted in \cite{ICIP2022} simultaneously, revealing very similar evolution. The plot also shows the 95\% confidence interval (CI) considering a Gaussian distribution, represented by the vertical bars for each bitrate, and by the green horizontal bar for the reference. It should be noted that, for most cases, MOS values obtained using the HMD are usually lower than those obtained using a 2D display for the same content. This suggests that the subjects are more sensitive to artifacts created by the codecs, even at higher bitrates. In the HMD test, some content did not show the typical monotonic behavior, observed in the 2D and 3D evaluations. This is the case for rates R02 and R03 of \textit{Romanoillamp} encoded with G-PCC, and rates R03 and R04 of \textit{Rhetorician} encoded with RS-DLPCC, where the MOS is slightly lower for the higher bitrates.
The reasons for this are difficult to explain for the authors. The quality is very similar and there might exist a particular reason that lead some naive subjects to give lower scores, which cannot be understood.

Table \ref{table:testCorrelation} shows the Pearson Correlation Coefficient (PCC), the Spearman Rank Order Correlation Coefficient (SROCC), the Root-Mean Squared Error (RMSE), and the Outlier Ratio (OR), between the MOS of previously conducted evaluations and the HMD experiment. The latter reveals high correlation values with both the 2D~\cite{EI2022} and 3D~\cite{ICIP2022} experiments. Moreover, a Kruskal-Wallis one-way analysis followed by a multiple comparison test was performed~\cite{KruskalWallis} (Table \ref{tab:statDif}), revealing no statistical differences between evaluations (\textit{p} $>$ 0.05).

Table~\ref{table:confidenceIntervals} shows the comparison between the average of the 95\% CIs of the subjective scores obtained using the HMD, a 3D display~\cite{ICIP2022}, and a regular 2D display~\cite{EI2022}. The average was computed across all source content for each codec-bitrate pair. In the \textit{Global} column, the global average of the 95\% CIs is shown.  As the number of scores influence the CI, only the first 16 subjects were considered to compute the CIs, when comparing with the 2D evaluation. For the comparison with the 3D evaluation, all 18 subjects were considered.
The CIs are significantly narrower for the evaluation conducted with the HMD, which indicates that subjects tend to agree more on the scores for each test instance.
Furthermore, the hidden references were always scored with 5, which was not the case in the experiments with the 2D and 3D displays (the reference MOS 95\% CI is represented with a green horizontal bar in Fig. \ref{MOSBPP}). 

\subsection{Objective Evaluation Results}
This study was complemented with an objective quality evaluation, using five full-reference point cloud quality metrics, notably PSNR MSE D1 (point-to-point mean square error), PSNR MSE D2 (point-to-plane mean square error)~\cite{8296925}, PointSSIM~\cite{9106005}, PCQM~\cite{9123147}, and GraphSIM~\cite{QiYangGraphSIM2022}. Their correlation with the MOS was assessed by the PCC, SROCC, RMSE, and OR, as recommended in~\cite{ITU-T}, are represented in Table \ref{table:metricsResult}. 
The predicted MOS were computed after logistic regression on the objective scores, as it is commonly done when benchmarking objective quality metrics\cite{HDRMarco}. Fig. \ref{fig:LogisticF} shows the normalized objective metric vs. normalized MOS plots for the two best performing metrics, PCQM and PSNR MSE D1. 
Although the correlations are generally slightly lower using the HMD, they are very similar to those of the 3D study, revealing these metrics can predict the compression quality in a similar fashion, compared with typical 2D tests.
PCQM yielded the highest correlation values for both evaluations, achieving a slightly better performance with the HMD.
\vspace{-0.4cm}
\section{Conclusions}
A quality subjective evaluation of point cloud codecs using a HMD was reported.
The obtained results suggest that conducting subjective evaluation using a HMD produces similar results, in comparison with 3D stereoscopic representation~\cite{ICIP2022} and 2D visualization~\cite{EI2022}. 
The MOS CIs were narrower in this experiment, mainly for point clouds where it is easier to define the quality degree, such as the reference point clouds, which were always scored with a 5. Moreover, producing the test was much easier, as no computationally expensive video generation was needed and no large uncompressed videos needed to be stored. 

In general, subjects evaluated the point clouds with lower scores when using the HMD, because it is very easy to understand differences between the reference and the distorted point clouds. 
\bibliographystyle{IEEEbib}
\bibliography{refs}

\end{document}